\documentclass[aps,pra,reprint,groupedaddress]{revtex4-1}
\usepackage{graphicx}

\begin{document}

% Use the \preprint command to place your local institutional report
% number in the upper righthand corner of the title page in preprint mode.
% Multiple \preprint commands are allowed.
% Use the 'preprintnumbers' class option to override journal defaults
% to display numbers if necessary
%\preprint{}

\title{The effect of self- and cross-phase modulation on photon-pairs generated by spontaneous four-wave mixing in integrated optical waveguides}

\author{Gary F. Sinclair}
\email[]{gary.f.sinclair@bristol.ac.uk}
\author{Mark G. Thompson}
\affiliation{Centre for Quantum Photonics, School of Physics, H. H. Wills Physics Laboratory, University of Bristol, Tyndall Avenue, Bristol BS8 1TL, United Kingdom}

\date{\today}

\begin{abstract}
A novel time-domain approach using the momentum operator is used to model spontaneous four-wave mixing in a lossless nonlinear waveguide.  The effects of self- and cross-phase modulation on the photon-pair production rate and heralded photon purity are investigated.  We show that in the special case where only one half of the photon-pair state is filtered that the generation rate and purity of the heralded photons are unmodified by the presence of self- and cross-phase modulation.  The significance of this special case arises when we consider heralded single-photon sources, where future schemes are likely to only filter the herald photon to ensure a high heralding efficiency is maintained.
\end{abstract}

% insert suggested PACS numbers in braces on next line
% insert suggested keywords - APS authors don't need to do this
%\keywords{}

%\maketitle must follow title, authors, abstract, \pacs, and \keywords
\maketitle

% body of paper here - Use proper section commands
% References should be done using the \cite, \ref, and \label commands
\section{Introduction}
A high brightness source of pure and indistinguishable single photons is a crucial building block for several quantum technologies, including quantum networks \cite{o2009photonic}, quantum computing and simulation \cite{PhysRevLett.100.060502, aspuru2012photonic} and quantum-enhanced sensing \cite{higgins2007entanglement}.  Indeed, if quantum computing is to exceed classical capabilities, a source of on-demand pure single photons is one of the key requirements \cite{PhysRevA.66.053805, PhysRevX.5.041007}.  Recently, significant progress on single-photon sources has been made across several platforms including quantum dots \cite{somaschi2016near}, nitrogen-vacancy colour centers \cite{babinec2010diamond} and parameteric sources such as microstructured optical fibers \cite{Dyer:08}, silica \cite{Spring:13} and silicon waveguides \cite{silverstone2014chip, PhysRevX.4.041047}.

Here, we examine the production of photon pairs by spontaneous four-wave mixing (SFWM) in a nonlinear waveguide \cite{Lin:06, Sharping:06}.   In the simplest case, a single photon-pair state can be used to exhibit nonclassical interference between two identical sources \cite{silverstone2014chip, silverstone2015qubit}.  Alternatively, one half of the pair can be used as a herald for the photon in the counterpart mode.  In this way, nonclassical interference can be demonstrated between photons generated in multiple different sources \cite{1367-2630-13-6-065005, technologies4030025, xiong2016active, PhysRevLett.100.133601, PhysRevLett.102.123603}, which offers a route towards scaling-up to higher photon-number experiments.  This passively heralded approach on its own does not scale well however, since the probability of multiple photons being generated simultaneously from independent sources drops off rapidly as the desired number of photons increases.  Although the photon pairs are generated non-derministically, proposals exist that allow many sources to be multiplexed together into a near-deterministic single photon source \cite{PhysRevA.66.053805, Mendoza:16}.  In addition to the technical challenges that high-speed, low-loss multiplexing will entail, it is also necessary to ensure that the process of detecting the herald photon will  not degrade the purity of the heralded counterpart.  We expect that this could be the case, since photon pairs generated by a parametric processes will exhibit correlations due to the requirement for energy and momentum conservation \cite{agrawal2007nonlinear, PhysRevA.64.063815}.  It is the detection of the herald photon, which will cause the remaining photon to be projected into a mixed state, depending on the degree of correlation between both halves of the generated photon pair.  However, the purity of the heralded photon is essential since it determines the visibility of nonclassical interference between identical sources \cite{1367-2630-13-6-065005}.  

Several techniques to reduce spectral correlations of the photon pair have been investigated.  Typically, correlations are reduced by careful consideration of the pump bandwidth and dispersion properties of the nonlinear material.  That is, by carefully balancing the requirements for energy and momentum conservation, which both impose spectral correlations, the combination of these constraints can produce weakly correlated photon pairs, suitable for high-purity heralding \cite{PhysRevA.64.063815,PhysRevA.91.013819, PhysRevLett.100.133601,1367-2630-10-9-093011}.  One significant advantage of this technique, is that the absense of filtering allows every photon pair generated to be used, thus allowing a higher production rate.  However, in integrated optical devices we typically do not have the same freedom to chose materials, wavelengths and polarisations that are characteristic of bulk experiments.  Therefore, an alternative approach is required.  The most common approach is to apply tight spectral filtering to the generated photon pairs along with a spectrally broad pump pulse.  Although this allows the production of high-purity heralded photons, it will come at the expense of the heralding rate, since many of the generated photon pairs will now be rejected by the tight filtering.  To compensate for this and maintain a reasonable heralding rate, the use of intense pump pulses is required.  However, such short and intense pulses are necessarily going to be accompanied by other nonlinear effects, such as self- and cross-phase modulation, which will result in spectral changes to the propagating pump pulse and co-propagating photon pairs.  Previous work has suggested that such effects should be small, so long as we remain in the regime where multiple pair generation is negligable \cite{helt2013parasitic}, although a complete model has not yet been developed.  It is the aim of this work to develop a complete model including self- and cross-phase modulation, developing on previous work done on photon pair production in fiber \cite{PhysRevA.92.053849}, and to determine their impact on the photon pair production rate and purity of the heralded photons.

\section{Modelling photon-pair production in the time-domain}
\subsection{Derivation of the joint temporal amplitude}
The most natural framework to describe spatial propagation of an electromagnetic field is presented by the momentum operator \cite{PhysRevA.35.4661, PhysRevA.44.500}.   From the definition  $\hat{M} = -i \hbar \frac{\partial}{\partial z}$ we can write down the spatial evolution equation,
\begin{equation}
-i \hbar \frac{\partial \vert \psi \rangle}{\partial z} = \hat{M}(z) \vert \psi \rangle.
\label{eqn:meqn}
\end{equation}
This is the spatial analog of the more usual temporal evolution described by the Hamiltonian.  In our work, we consider a nonlinear material, in which a strong classical pulse propagates along with two quantum fields, named the signal and idler modes.  The full momentum operator for this is derived in the Appendix.  However,  in the interaction picture  the part responsible for spontaneous four-wave mixing is simply given by:
\begin{equation}
\label{mppp:m}
\hat{M}(z) = \hbar \gamma e^{ i \Delta \beta_0 z}\int A_p(z, t)^2 \hat{A}_i^\dag (z,t) \hat{A}^\dag_s(z,t) dt.
\end{equation}
Here, $A_p$ is the classical pump field and $\hat{A}_s$ and $\hat{A}_i$ are the position- and time-dependent operators describing the evolution of the signal and idler modes under the influence of self- and cross-phase modulation and $\gamma=\frac{\omega_0 n_2}{c A_{eff}}$ is the nonlinear parameter \cite{agrawal2007nonlinear, Lin:07}.  The phase matching of the SFWM process is described by $\Delta\beta_0 = 2 \beta_{p, 0} - \beta_{i, 0} - \beta_{s,0}$ \cite{agrawal2007nonlinear}, where the wavevector is series expanded around the carrier frequency for each of the three modes: $\beta_a(\omega) = \beta_{a, 0} + \beta_{a, 1}(\omega - \omega_{a, 0}) + \frac{1}{2}\beta_{a, 2}(\omega-\omega_{a, 0})^2 + \cdots$, where $a \in \{p, s, i\}$.  As an example of typical material parameters, for a silicon wire waveguide with a pump wavelength around 1550 nm we have $n_2\approx 6 \times 10^{-18}$ $\mbox{m}^2/\mbox{W}$ \cite{lin2007dispersion, bristow2007two} and $A_{\mbox{eff}}\approx 0.2$ $\mu \text{m}^2$.  The time integral is taken over the entire domain $t \in (-\infty, \infty)$, as it is in subsequent intergrals, unless stated otherwise.

By expanding the solution of (\ref{eqn:meqn}) to first-order, that is, considering only a single pair of generated photons $\vert \psi (L) \rangle \approx \vert 0 \rangle + \vert \psi_{1,1} \rangle + \cdots$ , we can see that the photon-pair state at the output of the waveguide is \cite{PhysRevA.92.053849},
\begin{equation}
\label{mppp:os}
\vert \psi_{1,1} \rangle = i \gamma \int d z \int dt e^{i \Delta \beta_0 z} A^2_p(z,t) \hat{A}^\dag_i (z, t) \hat{A}^\dag_s (z, t) \vert 0 \rangle,
\end{equation}
where the spatial integral is taken over $z \in [0, L]$, where $L$ is the length of the waveguide.  In the interaction picture, we know that the signal and idler modes will evolve under the action of a lossless cross-phase modulation according to,
\begin{equation}
\hat{A}_a (z,t) = \exp \left [ i \theta_a (z,t) \right ] \hat{A}_a(0, t-\beta_{a, 1} z), \quad a=\{s, i\}
\end{equation}
where $\theta_a(z,t)$ is the phase shift of the signal or idler mode as a function of the position and time, and $\beta_{a,1} = 1/v_{g, a}$ is the first-order dispersion parameter, which is simply the reciprocal of the group velocity for that mode.  Similarly, the pump evolves under the action of self-phase modulation (and possibly some nonlinear absorption):
\begin{equation}
A_p(z,t) =  \exp \left [ i \theta_p(z,t) \right ] \vert A_p(z, t) \vert.
\end{equation}
In contrast to \cite{PhysRevA.92.053849} where a non-negligable group velocity dispersion (GVD) in optical fiber is assumed to result in walk-off between the generated signal and idler photons, we assume that all three optical modes share a common group velocity ($\beta_{p,1}, \beta_{s, 1},  \beta_{i, 1} \rightarrow \beta_1=1/v_g$).   This is because in our work we consider near-degenerate four-wave mixing in integrated optical waveguides \cite{Sharping:06}, where the detuning between signal, idler and pump photons is much smaller and the propagation length significantly shorter than the walk-off length between the pulses \cite{agrawal2007nonlinear}.  In this situation, it is much more realistic to neglect group-velocity dispersion and suppose all three fields propagate at the same group velocity.  

Given the state generated at the end of the waveguide (\ref{mppp:os}), we can define a joint probability distribution, the {\it joint temporal amplitude} (JTA), that describes the joint probability of detecting signal and ilder photons at time $t_s$ and $t_i$,
\begin{equation}
JTA(t_s, t_i) = \langle 0 \vert \hat{A}_i(L, t_s) \hat{A}_s (L, t_i) \vert \psi (L) \rangle.
\end{equation}
The form of the momentum operator (\ref{mppp:m}) ensures that photons are always generated simultaneously.  This, along with the absence of GVD, results in the JTA taking on a diagonal form where $t_p=t_s=t_i$, is the time at which the pair of signal and idler photons simultaneously arrive at the end of the waveguide.  Then we can write the JTA in terms of a single time parameter such that, $JTA(t_s, t_i) = JTA(t_s) \delta(t_s-t_i)$.  Using the relationships for the evolution of the pump, signal and idler modes we find that,
\begin{equation}
JTA(t_p) = i \gamma \int dz P \left [ z, t_c(z, t_p) \right ] \exp \left [i \Theta(z,t_p) \right ].
\label{eqn:jta}
\end{equation}
Here, $t_c= t_p-\beta_1(L-z)$ is the time that the photon pair must have been generated, given that it arrived at the end of the waveguide at time $t_p$ and $P(z,t)=\vert A_p(z,t) \vert^2$ is the pump power in Watts.  Clearly the photon pairs are generated in a superposition of positions over the length of the waveguide, proportional to the power of the pump at that point in time and space.  The phase term $\Theta(z, t_p)$ includes the effects of SPM on the pump up until the moment the photon pair is generated, and subsequent XPM from the generation point to the end of the waveguide, in addition to the usual phase matching term:
\begin{equation}
\begin{array}{rcl}
\Theta(z, t_p) &=& \Delta\beta_0z + 2 \theta_p \left ( z, t_c \right)  \\
&& + \theta_i(L, t_p) - \theta_i \left (z, t_c) \right )  \\
&& + \theta_s(L, t_p) - \theta_s \left (z, t_c \right ) .
\end{array}
\end{equation}
Typically, the evolution of the pump will be described by the nonlinear Schr\"{o}dinger equation \cite{agrawal2007nonlinear}, which can be used to model all of the non-linear effects that the strong classical pump pulse will experience.  A common material platform employed for integrated photon-pair production is silicon \cite{1367-2630-13-6-065005, silverstone2014chip}, which exhibits a refractive and absorptive Kerr nonlinearitiy, along with associated free carrier effects \cite{Yin:07}.  When the pump pulse is of sufficiently short duration and low intensity, then free-carrier effects can be neglected.  Following \cite{Yin:07} we can ensure that we are working in this regime if $\frac{h \nu}{\sigma_{\text{FCA}} T_0} \gg I_0$, where $h \nu$ is the energy of a pump photon, $\sigma_{\text{FCA}}$ is the free-carrier absorption coefficient, $T_0$ is the pulse duration and $I_0$ is the peak pulse intensity in units of W/m$^2$.  If we continue to neglect group-velcocity dispersion and free-carrier effects then we can transform into a retarded reference frame ($\tau = t - \beta_1 z$), where the evolution of the pump can be solved analytically \cite{Yin:07}.  To express the pump envelope in the retarded frame we say $P_{\text{ret}}(z, \tau) = P(z, \tau+\beta_1 z)$.  However, for convenience we will simply drop the subscript and the presence of the retarded time variable $\tau$ will be taken to imply that whatever function we are working with is expressed with respect to a retarded frame.  Then, the pump pulse dynamics are given by,
\begin{eqnarray}
P(z, \tau) &=& \frac{P(0, \tau) \exp \left (-\alpha z \right )}{1 + \alpha_2 P(0, \tau) z}, \label{eqn:power} \\
\theta_p(z, \tau) &=& \frac{\gamma}{\alpha_2} \ln \left [1 + \alpha_2 P(0, \tau) Z_{eff}(z) \right ], \label{eqn:phi} 
\end{eqnarray}
where $\alpha_2$ is the two-photon absorption coefficient \cite{lin2007dispersion, bristow2007two}, $Z_{eff}(z) = \left (1 - \exp[-\alpha z]\right)/\alpha$ is the effective length and $\alpha$ is the linear propagation loss.  Noting that  cross-phase modulation is always twice as strong as the self-phase modulation ($\theta_p(z,\tau) = \theta(z, \tau), \theta_{s \vert i}(z, \tau) = 2 \theta(z, \tau)$ we find that the JTA can be expressed as:
\begin{eqnarray}
\label{eqn:JTAgeneral}
JTA(\tau) &=& i \gamma \exp \left [ 4 i \theta(L, \tau) \right ] \\
&& \times \int P(z, \tau) \exp \left [ i \Delta \beta_0 z - 2 i \theta(z, \tau) \right ] dz \nonumber.
\end{eqnarray}
Again, we have made use of the diagonal form of the JTA in the retarded frame ($JTA(\tau_s, \tau_i) = JTA(\tau_s) \delta(\tau_s-\tau_i)$) to express the JTA in a more compact form.  Using the solutions presented above for the pump evolution, (\ref{eqn:power}) and (\ref{eqn:phi}), we can evaluate the integral for the JTA numerically if desired.  However, for our further work in this paper, we will proceed with the assumption that linear and nonlinear losses can also be neglected as this will expedite the devepment of simple analytical models and best convey the underlying physics of the SFWM process.  In any real experiment losses will primarily reduce the pair production rate and heralding efficiency, but will also complicate the form of the JTA to some extent.  However, unlike SPM and XPM, loss does not constitute a fundamental constraint on pair production and could be reduced to a negligable level by improvements in fabrication and materials.  Neglecting losses, the nonlinear phase shift is consequently given by $\theta(z,t) = \gamma P(0,\tau) z$, allowing us to evaluate the integral in (\ref{eqn:JTAgeneral}) to find an explicit form for the JTA:
\begin{eqnarray}
JTA(\tau) &=& i \gamma P(0,\tau) L \exp \left [ 3 i \gamma P(0,\tau) + \frac{i \Delta \beta_o L}{2} \right ] \nonumber \\ 
&& \times \mbox{sinc} \left [ \frac{\left [ \Delta\beta_0 - 2 \gamma P(0,\tau) \right ] L}{2} \right ]. 
\end{eqnarray}
Before we proceed, we make one final simplification to our model.  Firstly, we assume that given a short waveguide, the signal and idler modes which we are examining are always well within the phase-matching bandwidth ($\Delta \beta_0 L \approx 0$).  This is a reasonable assumption, as bandwidths are typically several THz (e.g. for the 3 mm silicon waveguide used in \cite{technologies4030025} with dimensions 220 $\times$ 460 nm the bandwidth was 6 THz).  Secondly, the peak pulse power will cause small shifts in position of the phase-matching band, which are assumed to be negligable compared to the bandwidth itself.  These approximations are quite reasonable in the quite short integrated optical waveguides that are typically used.  With these two assumptions, the JTA takes on a particularly simple and intuitive form,
\begin{equation}
JTA(\tau) = i \gamma P(0,\tau) L \exp \left [3 i \gamma P(0,\tau) L \right ].
\label{eqn:jtasxpm}
\end{equation}
In words, the probability of generating a photon pair at time $\tau$ is simply proportional to the pump power at that time, and the phase-shift experienced by the photon pair is given by the average of the effects of self- and cross-phase modulation over the length of the waveguide.  We note, that the JTA given above describes the simultaneous arrival of photons at the waveguide exit, but should more fully be written as:
\begin{equation}
JTA(\tau_s, \tau_i) = JTA(\tau_s) \delta(\tau_s - \tau_i)
\label{eqn:JTAdiag}
\end{equation}
To reiterate, the simultaneous arrival of photons at the waveguide exit is due to the simultaneous generation of the photons described by (\ref{mppp:m}) and the absense of walk-off between the signal and idler photons, since we have neglected group velocity dispersion.

\subsection{Include filtering of the signal and idler photons}
In our analysis we have used a spatio-temporal model of SFWM to include the effects of self- and cross-phase modulation, since these lend themselves more readily to a description in this domain.  However, in the second part of our analysis we wish to include filtering of the signal and idler modes, which is commonly employed to improve the separability of the two-photon state.  Naturally, this is more readily described in the frequency domain.  Thus, given the JTA, we have two options with how we can proceed: either we remain in the time-domain or we use a Fourier transform to convert the joint-temporal amplitude into the more usual joint-spectral amplitude (JSA).  The convenience of either approach will depend on the particular problem to be solved, although both must necessarily furnish the same results.  Below, we choose the former option and remain in the time-domain, since this will prove the more insightful approach later, when we consider filtering of one mode only.

In the section above, it was assumed that photon pairs generated simultaneously in the waveguide will also be detected simultaneously at the output, if we neglect walk-off.  However, in practice the finite bandwidth of the photon detection process will fundamentally limit the precision with which we can specify the arrival time of both photons.  Typically, narrow-bandwidth filtering ($\approx 25 - 200$ GHz) is used to ensure high purity of the generated photons when using a ps-duration pulsed pump at telecom wavelengths.   In this case, the finite detection bandwidth will lead to a significant broadening of the otherwise diagonal JTA.  In addition, it could be expected that the temporal resolution of the detector might also increase the uncertainty of the photon arrival time.  The degree of temporal uncertainty introduced will depend on the lifetime of the coherent evolution of the detector, before decoherence results in a classical detection event.  This will likely be much less than the experimentally observed temporal resolution of the detector that will be dominated by sources of classical noise.  For a typical superconducting nanowire single-photon detector we expect the coherent timescale to be no greater than the lifteime of the excited electron, estimated to be about 2 ps \cite{engel2015detection}.  Therefore, in most current experiments we expect that the narrow-bandwidth filtering will dominate the blurring of temporal correlations.  Making this assumption, the filtered JSA is given by,
\begin{equation}
JSA_f(\Delta_s, \Delta_i) = JSA(\Delta_s, \Delta_i) f_s(\Delta_s) f_i(\Delta_i),
\end{equation}
where $f_a(\Delta_a)$ is the field amplitude filter function and $\Delta_a = \omega_a - \omega_{a,0}$ is the detuning of the field from the carrier frequency for the signal or idler modes  ($a \in \{s, i \}$).   As filtering in the frequency domain is always simply the product of the filter function and the input, in the time-domain this becomes a convolution:
\begin{eqnarray}
JTA_f(\tau_s, \tau_i) &=& \frac{1}{2 \pi} \int \int  JTA(\tau_s', \tau_i') \nonumber \\ 
&&  \times f_s(\tau_s - \tau_s') f_i(\tau_i -\tau_i') d\tau_s' d\tau_i'.
\label{eqn:JTAdouble}
\end{eqnarray}
Here, the time-domain filter function $f_a(\tau_a)$ is the Fourier transform of the frequency-domain function $f_a(\Delta_a)$ on mode $a \in \{s, i\}$.  We can see that the effect of the finite filter bandwidth, is to blur the arrival times of the photons in the signal and idler modes, according to the point-spread functions $f_a(\tau_a)$.  In the section above, we saw that the unfiltered JTA was diagonal, that is, both photons will arrive simultaneously at the output of the waveguide.  We can use this to simplify the double-integral in (\ref{eqn:JTAdouble}) by rotating into a set of axes $\tau_\alpha = (\tau_s + \tau_i)/\sqrt{2}$ and $\tau_\beta = (\tau_i - \tau_s)/\sqrt{2}$.  Then by virtue of the Dirac delta function in (\ref{eqn:JTAdiag}), we reduce the expression to:
\begin{equation}
JTA_f(\tau_\alpha, \tau_\beta) = \frac{1}{2 \pi} \int \frac{dT}{\sqrt{2}} JTA \left ( \frac{T}{\sqrt{2}} \right ) g(\tau_\alpha - T, \tau_\beta),
\label{eqn:jtaf}
\end{equation}
where $g(x, y) = f_s \left ( \frac{x-y}{\sqrt{2}} \right ) f_i \left ( \frac{x + y}{\sqrt{2}} \right )$ is the filter function rotated into the new co-ordinate basis and $JTA(\tau)$ is the unfiltered diagonal JTA defined by (\ref{eqn:JTAdiag}).  What this decomposition represents is the broadening of the original diagonal JTA by the filter functions $g(\tau_\alpha, \tau_\beta)$.  The originally precisely defined and simultaneous arrival times of the signal and idler photons are blurred, due to the finite filtering bandwidth.  The filter functions themselves, can be viewed as forming an over-complete basis for the JTA.

Since the JTA represents the joint probability of detecting a pair of photons, we can write the output state of the waveguide, including filtering, in terms of the JTA.  The effect of filtering is to probabilistically remove signal and idler photons.  This will result in an output state that is a mixture of photon number states $\hat{\rho} = \hat{\rho}_{00}+\hat{\rho}_{10} + \hat{\rho}_{01} + \hat{\rho}_{11}$, where the term $\hat{\rho}_{xy}$ represents a component with $x$ signal and $y$ idler photons.  Considering only the pair-photon term, for which we can define a joint probability distribution, we can say $\hat{\rho}_{11} = \vert \psi_{1,1} \rangle \langle \psi_{1,1} \vert$, where,
\begin{equation}
\label{fsi:psi}
\vert \psi_{1,1} \rangle = \frac{1}{2 \sqrt{2 \eta} \pi} \int JTA \left ( \frac{T}{\sqrt{2}} \right ) \vert \phi(T) \rangle dT.
\end{equation}
Here, $\eta$ is a normalisation constant, equal to the probability of detecting a single photon-pair at the output of the waveguide, and the $\vert \phi(T) \rangle$ is a family of factorisable, although over-complete, basis states.  These basis state are defined by the filter functions:
\begin{eqnarray}
\vert \phi (T) \rangle &=& \int \int f_s \left ( \tau_s - \frac{T}{\sqrt{2}} \right ) f_i \left ( \tau_i - \frac{T}{\sqrt{2}} \right )  \nonumber \\ 
&& \times \vert 1_{\tau_s} \rangle \otimes \vert 1_{\tau_i} \rangle  d \tau_s d \tau_i.
\label{eqn:ffb}
\end{eqnarray}
Each state $\vert \phi (T) \rangle$ represents the most precise statement we can make about the temporal mode of a photon-pair given the finite bandwidth of the signal and idler modes.  The overlap between the filter function basis states are given by the product of the overlaps of the signal and idler parts independently, $ \langle \phi (T') \vert \phi (T) \rangle = O_s(T-T') O_i(T- T')$, where $O_a(T-T')=\int f_a (\tau_a - T/\sqrt{2} )f_a (\tau_a - T'/\sqrt{2} )d\tau_a$ is the overlap of the signal (or idler) filter funtion at different times.  

Using the form of the output state given above in (\ref{fsi:psi}) the probability of generating a photon pair is given by:
\begin{eqnarray}
\label{fsi:eta}
\eta &=&\frac{1}{8 \pi^2} \int \int JTA \left ( \frac{T'}{\sqrt{2}} \right ) JTA \left ( \frac{T}{\sqrt{2}} \right ) \nonumber  \\
&& \times O_s(T-T') O_i(T-T')  dT' dT.
\label{eqn:eta}
\end{eqnarray}
Typically, photon pair sources based on parametric processes such as SFWM are used in the regime where $\eta \approx 0.1$ to avoid contamination of the output state by higher-order photon-pair terms, as this is highly detrimental to the visibility of non-classical interference \cite{1367-2630-13-6-065005}.  Due to the truncation of the output state to single photon-pair generation events, the effects of higher-order terms are not studied in this work.  Therefore it is worth noting that the range of validity for our work is for values where the integral in (\ref{fsi:eta}) evaluates to a photon-pair production probability of $\eta \leq 0.1$.  This integral in turn then determines the limits of the pump power, pulse duration and filter bandwidths which we can use and nonetheless remain in the low-excitation limit that we require to achieve high-visibility non-classical interference.

Of critical importance to the quality of a heralded single photon source, is the purity of the single photons produced.  If we post-select on the detection of a photon-pair, then we can write the density matrix of the two-photon state as $\rho = \vert \psi_{1,1} \rangle \langle \psi_{1,1} \vert$ and note that the purity of the heradled single photon is given by $\mathcal{P} = tr_i (\hat{\rho}_i^2)$ where $\hat{\rho}_i = tr_s({\hat{\rho}})$ is the state vector after we have traced-out, or detected, the signal (herald) photon.  This gives us the expression for the purity,
\begin{eqnarray}
\mathcal{P} &=& \left ( \frac{1}{2 \sqrt{2 \eta} \pi}\right )^4  \int  JTA \left ( \frac{T_1}{\sqrt{2}} \right ) JTA^* \left( \frac{T_2}{\sqrt{2}} \right ) \nonumber \\
&& \times JTA \left (\frac{T_3}{\sqrt{2}} \right ) JTA^* \left (\frac{T_4}{\sqrt{2}} \right )  O_s(T_1-T_2) \nonumber \\
&& \times  O_s(T_3-T_4) O_i(T_1-T_4) O_i(T_3-T_2) d^4 T.
\label{eqn:p}
\end{eqnarray}

\begin{figure*}[t]
\includegraphics[width=1.9 \columnwidth]{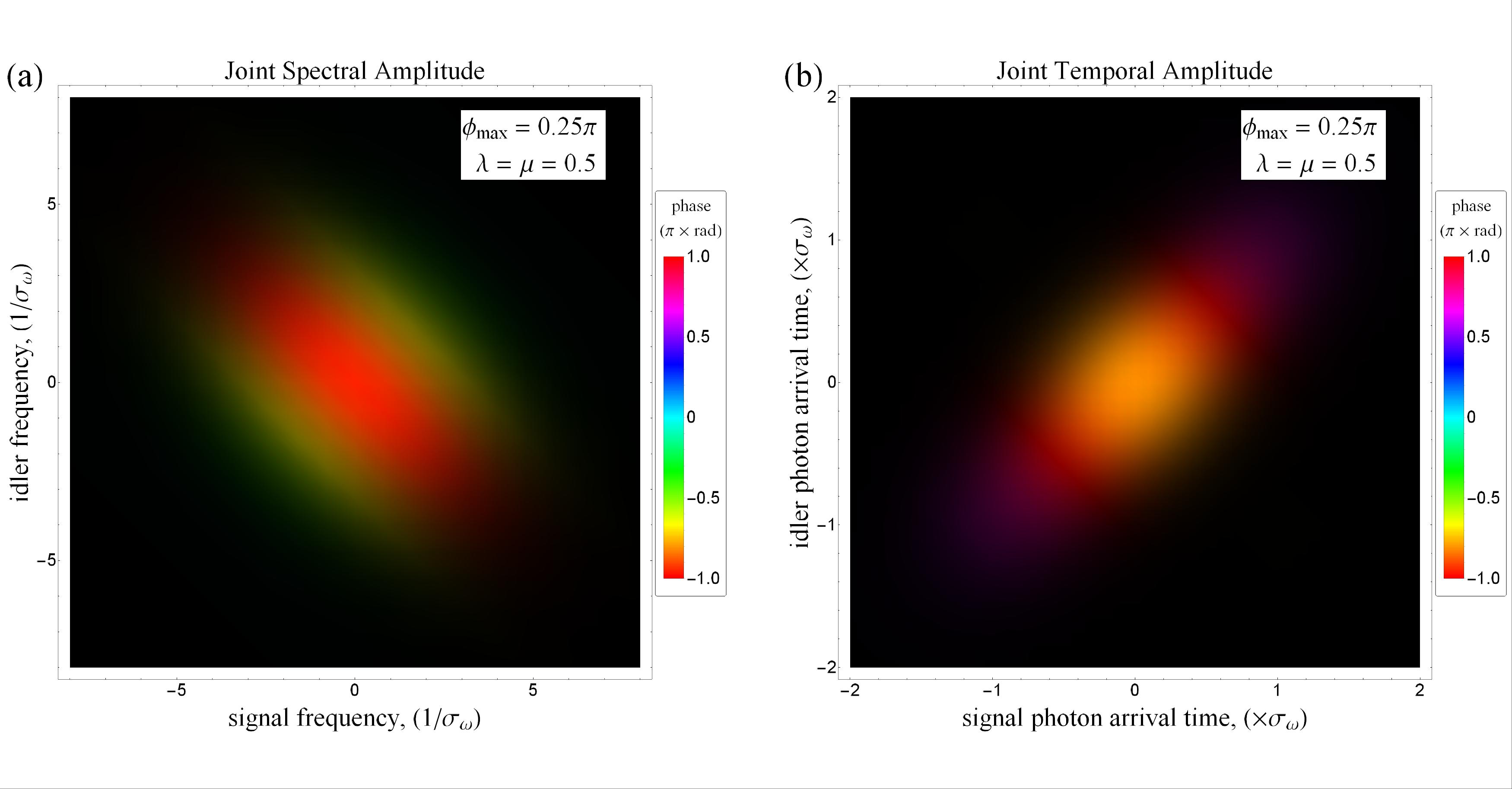}
\caption{\label{fig:jsajta}The joint spectral amplitude (JSA) and joint temporal amplitude (JTA)  produced by assuming a Gaussian pump profile and Gaussian filters on the signal and idler modes.  The colour of the plots represents the phase imparted to the state by self- and cross-phase modulation and the brightness represents the detection probability.  In this example, the filtering is not very tight, and the alignment of the joint spectrum (a) along the energy conservation line (top-left to bottom-right) is clear.  Similarly, in (b) the alignment of the JTA along the line of  simultaneous photon arrival (bottom-left to top-right) is also easily distinguishable.  Both plots are related to each other by a Fourier transform.}
\end{figure*}

\section{Examples}
\subsection{No self- or cross-phase modulation with Gaussian pump and filters}
The purpose of developing the spatio-temporal model above was to a construct a convenient framework to understand the effects of self- and cross-phase modulation on photon-pair production.  However, in this section we begin by considering the situation where the nonlinear phase shift is negligable; an approximation appropriate at low pump powers.  This provides the simplest example of applying the methods presented, and furnishes useful results for comparison with the full model developed later.  In the situation where the nonlinear phase shift is small, we can approximate the unfiltered JTA as,
\begin{equation}
JTA(\tau_s, \tau_i) = i \gamma P(0, \tau_s) L \delta (\tau_s-\tau_i).
\end{equation}
For the convenience of performing analytical calculations we assume that the filter and pulse profiles are both Gaussian.  Although in real experiments this may not be the closest representation of the real profiles, it provides a convenient approximation to the trends observed in real experiments.  We take the pump and filter profiles in the time domain to be:
\begin{eqnarray}
P(0, \tau) &=& P_0 \exp \left [\frac{-\tau^2}{2 \sigma^2_t} \right ], \\
f_a(\tau_a) &=& \sqrt{2} \sigma_{f, a} \exp \left [ - \sigma_{f,a}^2 \tau_a^2  \right ], \quad a \in \{s, i\}.
\end{eqnarray}
Here, $P(0, \tau)$ is the input temporal profile for a Gaussian pump pulse with width $\sigma_t = 1/ (2 \sigma_\omega)$  and $f(\tau)$ is the Fourier transform of a field filter profile, $f(\Delta_a) = \exp (-\Delta_a^2/(4 \sigma_{f, a}^2) )$.  Substituting these definitions for the pump and filter profiles into (\ref{eqn:jtaf}) we find the JTA including filtering is,
\begin{eqnarray}
JTA_{f}(\tau_s, \tau_i) &=& \frac{i \gamma L P_0}{\sqrt{\pi}} \sigma_\omega \frac{1}{\sqrt{2 \lambda^2 \mu^2 + \lambda^2 + \mu^2}} \\
&& \times \exp \left [ - \frac{\sigma_\omega^2 \left \{ 2 \mu^2 \tau_s^2 + 2 \lambda^2 \tau_i^2 + (\tau_i - \tau_s)^2 \right \}}{2 \lambda^2 \mu^2 + \lambda^2 + \mu^2} \right ], \nonumber
\end{eqnarray}
where $\lambda = \sigma_\omega / \sigma_{f, s}, \mu = \sigma_\omega/\sigma_{f,i} $ are the ratios of the pump to filter bandwidths for the signal and idler modes.  Using equation (\ref{eqn:p}) we find a particuarly simple expression for the the purity:
\begin{equation}
\mathcal{P} = \sqrt{1-\frac{1}{(1+2 \lambda^2)(1 + 2 \mu^2)}}.
\label{eqn:plin}
\end{equation}
As expected, we find that the purity of the heralded single photon tends towards unity as we reduce the bandwidth of either the signal or idler filter ($\lambda \rightarrow \infty$ or $\mu \rightarrow \infty$).
 Also, from equation (\ref{eqn:eta}) we can determine the probability of generating a photon pair.
\begin{eqnarray}
\eta(\lambda, \mu)  &=& \frac{(\gamma L P_0)^2}{2 \sqrt{2}} \frac{1}{\sqrt{\lambda^2 + \mu^2 + 2 \lambda^2 \mu^2}}, \\
&&=\frac{(\gamma L P_0)^2}{2} \sqrt{\frac{1-\mathcal{P}^2}{\mathcal{P}^2}}.
\label{eqn:etalin}
\end{eqnarray}
Again, this exhibits the expected quadratic dependence on pump power that we expect for a four-wave mixing process.  Interestingly, the assumption of Gaussian pump and filters also leads to a very simple trade-off between the pair production rate and purity.  That is, for a fixed pump power, an increase in purity (by tighter filtering) will necessarily decrease the pair production rate, as demonstrated by the remarkably simple relationship given in (\ref{eqn:etalin}).

Since we are filtering on both the signal and idler modes, the heralding efficiency will in general be less than one.  This is due to the fact that, on occassion, a hearld photon will be detected at the signal wavelength, whose counterpart (heralded) idler photon has been dropped by the filter and so is lost.  We define the heralding efficiency as the probability that a heralded photon will be detected, given that a herald has been registered and find that this can be calculated as the ratio of the pair-production rate with filtering at both signal and idler wavelengths to the pair production rate with filtering only on the signal: $\nu = \eta(\lambda, \mu) / \eta(\lambda, 0)$.
\begin{equation}
\nu = \sqrt{2} \lambda \sqrt{\frac{1-\mathcal{P}^2}{\mathcal{P}^2}}
\label{eqn:nu}
\end{equation}
For completeness, we note that these calcuations could have also been performed in the frequency domain.  Our first step, would have been to Fourier transform the JTA into the frequency domain to yield the JSA:
\begin{equation}
JSA(\Delta_s, \Delta_i) = i \gamma L \tilde{P}(\Delta_s+\Delta_i),
\end{equation}
where $\tilde{P}(\Delta) = \mbox{F.T.}\left [ P(\tau) \right]$ is the Fourier transform of the pulse profile in time, not the power spectrum of the pulse.  Since we earlier neglected the phase mismatch, this represents a SFWM process of infinite phase matching bandwidth.  In practice, we always apply some filtering to at least one of the generated signal and idler photons.  Assuming a filter bandwidth of $FWHM=2 \sqrt{2 \ln 2} \sigma_{f,a}$, where $a \in \{s, i\}$ for the signal and idler modes then we find the anticipated result:
\begin{eqnarray}
JSA_f(\Delta_s, \Delta_i) &=& \frac{i \gamma L P_0}{2} \frac{1}{\sqrt{2 \pi} \sigma_\omega} \exp \left [ - \frac{\left ( \frac{\Delta_s + \Delta_i}{2} \right )^2 }{2 \sigma_\omega^2} \right ] \nonumber \\ 
&& \times \exp \left [-\frac{\Delta_s^2}{4 \sigma_{f, s}^2} \right ] \exp \left [ - \frac{\Delta_i^2}{4 \sigma_{f, i}^2} \right ].
\end{eqnarray}
Noting that the representation of the two photon state in the frequency-domain is $\vert \psi_{1,1} \rangle = \frac{1}{\sqrt{\eta}} \int JSA(\Delta_s, \Delta_i) \vert 1_{\Delta_s} \rangle \otimes \vert 1_{\Delta_i} \rangle d\Delta_s d\Delta_i$, and $\hat{\rho}=\vert \psi_{1,1} \rangle \langle \psi_{1,1} \vert$ we find that the purity is exactly as was given before:
\begin{equation}
\mathcal{P} = \sqrt{1-\frac{1}{(1+2 \lambda^2)(1 + 2 \mu^2)}}.
\end{equation}
Naturally, other quantities such as the pair production rate and heralding efficiency can also be calculated in a similar manner to that undertaken in the time-domain, with identical results yielded.

\subsection{Include self- and cross-phase modulation with Gaussian pump and filters}
\begin{figure}[t]
\includegraphics[width=0.9 \columnwidth]{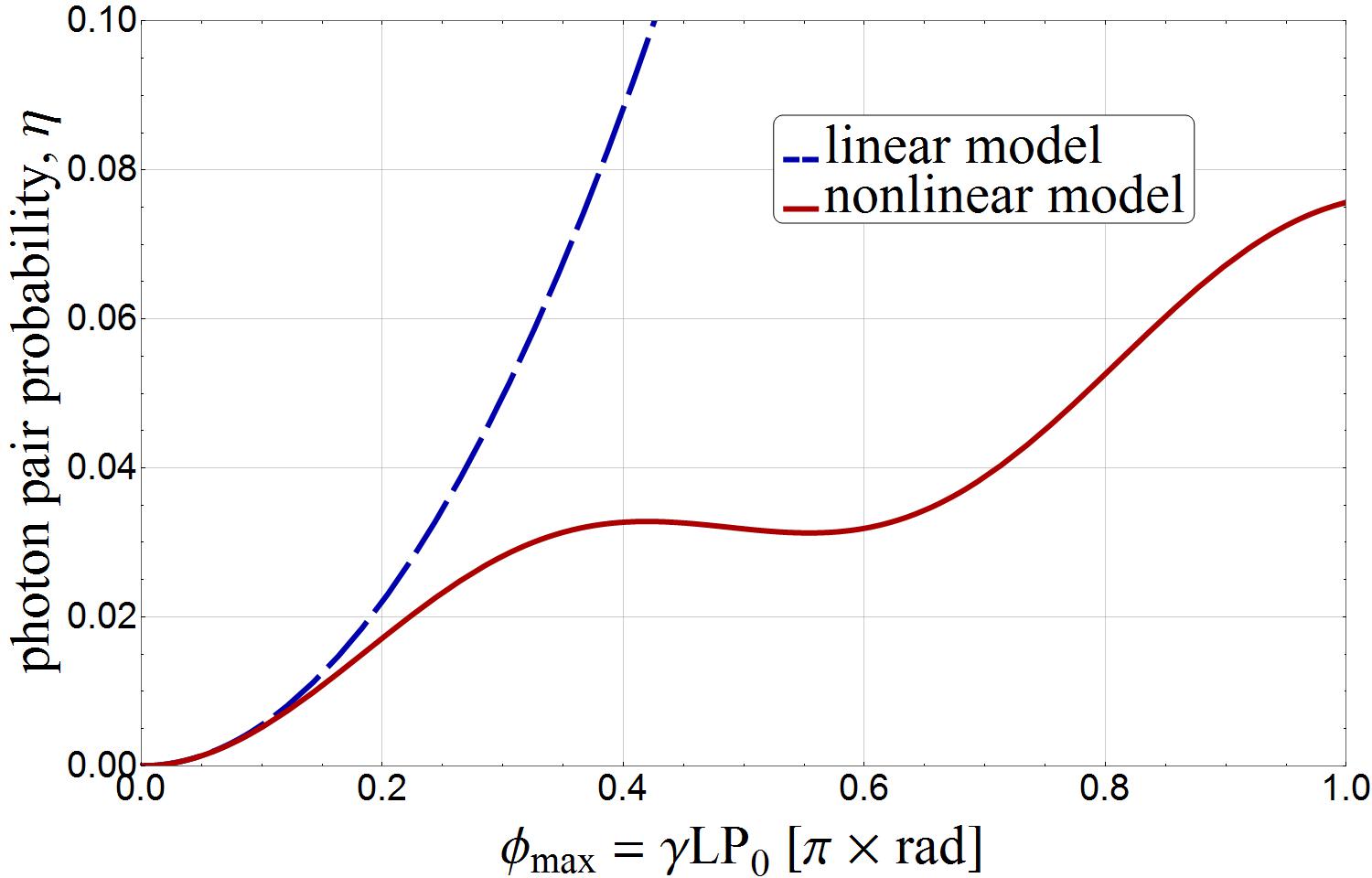}
\caption{\label{fig:eta}The photon pair production rate versus $\phi_{max}=\gamma L P_0$ (the nonlinear phase-shift of the pump pulse peak) for $\lambda = \mu = 2$.  The linear model exhibts the quadratic dependence on pump power expected.  However, when the effects of self- and cross-phase modulation are included the pair production probabiliy is significantly modified.  The step-wise increase in generation rate is due to both SPM and XPM spectrally broadening the generated photons and pushing them outside of the filtering bandwidth.}
\end{figure}

\begin{figure}[t]
\includegraphics[width=0.9 \columnwidth]{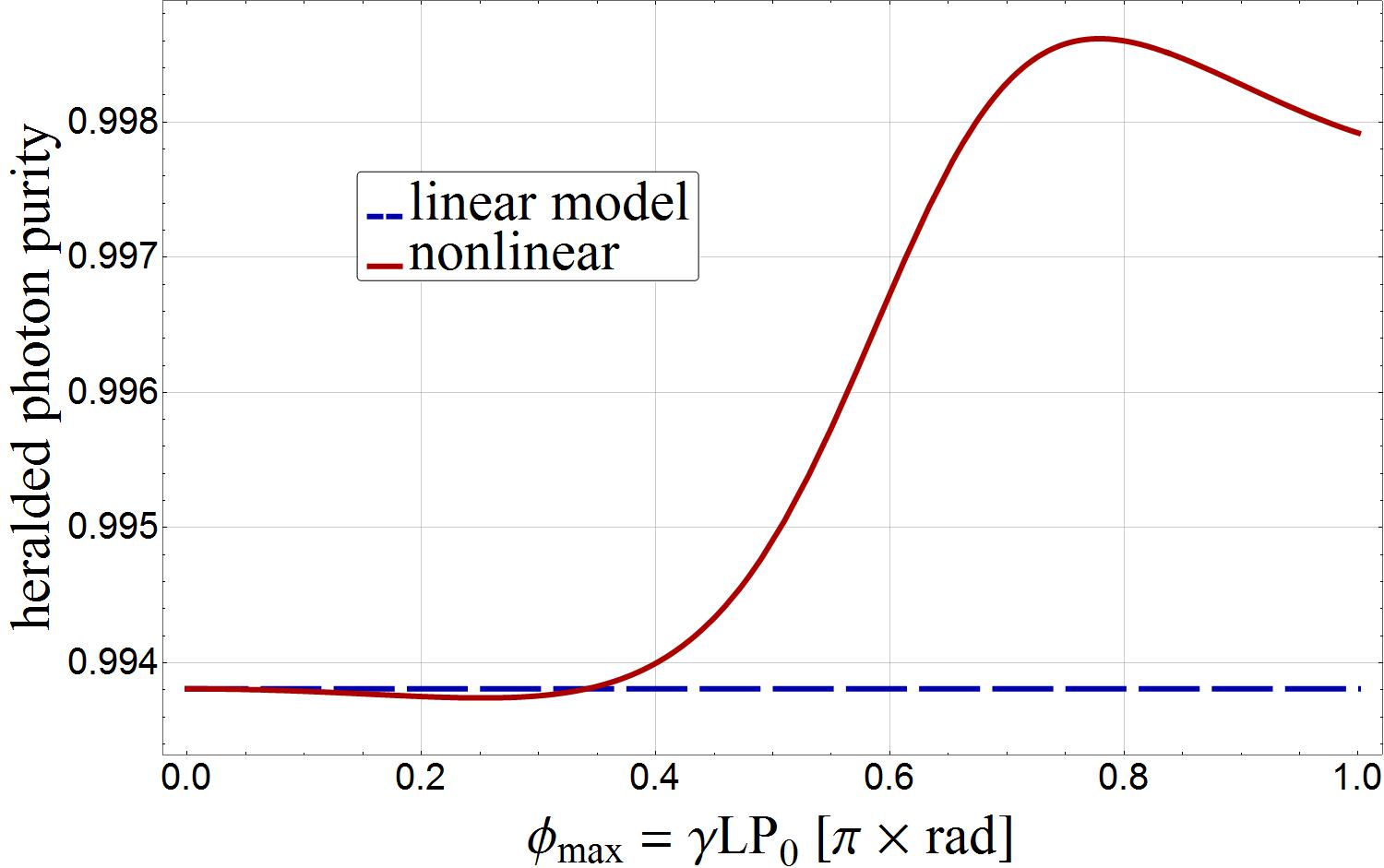}
\caption{\label{fig:p}The heralded photon photon purity versus $\phi_{max}=\gamma L P_0$ (the nonlinear phase-shift of the pump pulse peak) for $\lambda = \mu = 2$.  The linear model predicts a purity that is independent of the pump power.   However, when the  self- and cross-phase modulation are included the purity is shown to be slightly modified.  Spectral broadening of the pump and generated photon pairs provides a modest improvement to the purity of the heralded photon over most of the range of pump powers.}
\end{figure}

More generally, we should include the effects of self- and cross-phase modulation in our model of photon pair production.  Our departure point for this calculation is now given by (\ref{eqn:jtasxpm}), the JTA including the nonlinear phase shift.  Again, the JSA and JTA will remain Fourier transforms of each other, and any quantity that can be calculated in the time-domain can of course be calculated in the frequency domain.  However, given particular examples, it may sometimes be more convenient to remain in one domain than the other.  Working in the time-domain and assuming Gaussian filters and pump profile, using (\ref{eqn:jtaf}) we find that the JTA including filtering is given by,
\begin{eqnarray}
JTA_f(\tau_\alpha, \tau_\beta) &=& \frac{ i \phi_{max} \sigma_{f,s} \sigma_{f,i}}{ \sqrt{2} \pi} \sum^\infty_{n=0} \frac{(i 3 \phi_{max})^n}{n!} \nonumber \\
&& \times \int dT \exp \left [ - \frac{T^2(n+1)}{4 \sigma_t^2} \right ] \nonumber \\
&& \times \exp \left [ - \frac{\sigma_{f,s}^2}{2} (\tau_\alpha-T-\tau_\beta)^2 \right ] \nonumber \\ 
&& \times \exp \left [- \frac{\sigma_{f,i}^2}{2} (\tau_\alpha - T - \tau_\beta)^2  \right ],
\label{eqn:jtaf}
\end{eqnarray}
where we are again working in the rotated coordinate system  $\tau_\alpha = (\tau_s + \tau_i)/\sqrt{2}$ and $\tau_\beta = (\tau_i - \tau_s)/\sqrt{2}$ and $\phi_{\text{max}}=\gamma L P_0$ is now identified as the phase-shift of the peak of the pump pulse after propagating to the end of the waveguide.  It is worth commenting that the summation in (\ref{eqn:jtaf}) arises due to the series expansion of the nonlinear phase term in  (\ref{eqn:jtasxpm}).  Since each term in this summation is Gaussian, we can evaluate the integrals to give,
\begin{widetext}
\begin{equation}
\label{eqn:jtaexample}
JTA_f(\tau_s, \tau_i) = \frac{i \phi_{max}}{\sqrt{\pi}} \sigma_\omega \sum^\infty_{n=0} \frac{(i 3 \phi_{max})^n}{n! \sqrt{2(1+n) \lambda^2 \mu^2 + \lambda^2 + \mu^2}} \exp \left [- \frac{\left \{ 2(1+n) ( \lambda^2 \tau_i^2+\mu^2 \tau_s^2) + (\tau_i - \tau_s)^2  \right \} \sigma_\omega^2}{ 2(1+n) \lambda^2 \mu^2 + \lambda^2 + \mu^2} \right ].
\end{equation}
\end{widetext}
Again, $\lambda = \sigma_\omega / \sigma_{f, s}, \mu = \sigma_\omega/\sigma_{f,i}$ are the ratios of the pump to filter bandwidths for the signal and idler modes.  A plot the JTA given by (\ref{eqn:jtaexample}) and its Fourier transform, the JSA, is shown in Fig. \ref{fig:jsajta}.  With this solution for the JTA we can use equation (\ref{eqn:eta}) to calculate the photon pair production rate and (\ref{eqn:p}) to calculate the purity, although no convenient analytical forms exist for these quantities.  Nonetheless the relevant integrals can be calculated numerically and are plotted for some typical values in Fig. \ref{fig:eta} and Fig. \ref{fig:p}.  These both show notable departures from the predictions of the linear model.  In the case of the photon-pair production rate, we can see that when the filtering is quite tight (to achieve high purity) the production rate drops below that of the linear model.  This can be understood as due to the spectral broadening of both the pump and generated photon pairs that arises due to the additional nonlinear phase term in (\ref{eqn:jtasxpm}).  The effect of this is to broaden the spectrum of the generated photons, pushing them outside of the filtering bandwidth, and hence reducing the detection rate at the end of the waveguide.   Fig. \ref{fig:p} on the other hand, shows an improvement in the purity of the heralded photons.  Qualitatively, this is due to the flattening of the generated joint-spectrum, again due to spectral broadening, that in this particular case makes the state more factorisable.  For the calcuation of purity, rather than using (\ref{eqn:p}) that involves a four-fold integral, it is frequently easier to perfrom a numerical Schmidt (or singular-value) decomposition of the JTA to calculate the purity that way.  Although both approaches will naturally lead to the same values, several scientific computation packages offer convenient and efficient methods for undertaking singular-value decompositions.  When employing the Schmidt decomposition, we note that the purity is given by $\mathcal{P} =\sum \vert g_k \vert^4$, where $g_k$ is the weight of each Schmidt mode present in the decomposition of the JTA or JSA \cite{1367-2630-10-9-093011}.  It is interesting to note the similarity between (\ref{eqn:p}), which relates the JTA to the purity and the well-known Schmidt decomposition, mentioned above.  The main difference being the non-trival integral kernel in (\ref{eqn:p}) that arises due to the non-orthogonality of the filter function basis states defined by (\ref{eqn:ffb}), compared to the orthogonal Schmidt modes.

\subsection{Include self- and cross-phase modulation and filtering of the herald photons only}
Ideally we do not wish to filter the heralded photons as this will reduced the heralding efficiency, which is a key metric for a heralded single-photon source.  Here, we examine the behaviour of the heralded single-photon source, when only the signal (herald) photon is filtered.  In this way, we can control the purity of the source by adjusting the signal to pump bandwidth, while maintaining a theoretical heralding efficiency of unity.  Taking the limit where the idler filter bandwidth tends to infinity ($\sigma_{f, i} \rightarrow \infty$), we find that the filter in the time domain tends towards a Dirac delta function.  This implies that the unfiltered, and diagonal, JTA will not be broadened along the idler axis when performing (\ref{eqn:jtaf}) to calculate the filtered JTA.  Similarly, we find that the time-domain filter overlap function becomes,
\begin{equation}
\lim_{\sigma_{f,i} \to \infty} O_i(T-T') = 2 \sqrt{2} \pi \delta (T-T').
\end{equation}
Substituting this into (\ref{eqn:p}) for the purity we show that:
\begin{eqnarray}
\mathcal{P} &=& \frac{1}{(2 \sqrt{2} \pi \eta)^2}\int \int \left \vert JTA \left ( \frac{T}{\sqrt{2}} \right ) \right \vert^2  \nonumber \\
&& \times \left \vert JTA \left ( \frac{T'}{\sqrt{2}} \right ) \right \vert^2  \vert O_s(T-T') \vert^2.
\end{eqnarray}
The presence of the Dirac delta function in the integral has reduced (\ref{eqn:p}) to a two-fold integral, dependent only on the magnitude, and not the phase, of the JTA.  Since the only effect of self- and cross-phase modulation in the time domain is to introduce a nonlinear phase on the JTA. the purity reduces back to the simple linear model, as given by (\ref{eqn:plin}).  This important results is only apparent in the time-domain, since when viewed in the frequency domain the accumulated phase results in broadening of the pump pulse and of the generated photons.  Although these do still occur when filtering is applied to the herald photon only, resulting in a marked change to the JSA, they do not change the purity of the heralded photons.  When viewed in the frequency domain, changes in amplitude and phase of the JSA will, remarkably, cancel out leaving the weights of the Schmidt decomposition, $g_k$, unchanged.

Similarly, using equation (\ref{eqn:eta}) we find that the pair-production probability becomes:
\begin{equation}
\eta = \frac{O_s(0)}{2 \sqrt{2} \pi} \int \left \vert JTA \left ( \frac{T}{\sqrt{2}} \right ) \right \vert^2 dT.
\end{equation}
Again,  when evaluated this reduces back to the same expression as given by the simple linear model (\ref{eqn:etalin}), since the integral depends only on the magnitude, and not the phase, of the JTA.

\section{Conclusion}
Spontaneous four-wave mixing in integrated optical waveguides offers a promising route to realising  pure, indistinguishable and, if several such sources are multiplexed together, on-demand single photons.  A key requirement for parametric sources, is that the heralded single photon must remain pure when the herald is detected, despite the spectral correlations between signal and idler photons imposed by energy conservation.  The use of tight spectral filtering of the herald photon allows the effect of the spectral correlations to be reduced, without impacting upon the the heralding efficiency.  However, to  maintain a reasonable source brightness a strong pulsed pump is therefore required.  Naturally, the use of an intense pump pulse in a nonlinear medium will be accompanied by self- and cross-phase modulation, which will necessarily have an impact on the pump spectrum and joint spectrum of the generated photon pairs.  These are in turn expected, and do, affect the purity and photon pair production rate of the source.

In this work, we calculate the effects of self- and cross-phase modulation on the generated photon pairs using the momentum operator in a novel time-domain approach.  The time domain is shown to be the most natural framework for exploring the impact of nonlinearity on the generated photons.  Example analytical models are developed that show how the purity and photon pair production rate are in general modified by the nonlinearity.  However, in the particular case where only one half of the photon pair is filtered (usually the herald) then despite spectral changes to the JSA, remarkably, the effects of self- and cross-phase modulation vanish.  This is shown to be due to the independence of the photon purity on the phase of the JTA, through which the nonlinearity usually acts.  This result is of particular significance, since we are only likely to filter the herald photon in a multiplexed single photon source to maintain a high heralding efficiency.

\begin{acknowledgments}
This work was supported by the Engineering and Physical Sciences Research Council (EPSRC) and The European Research Council (ERC).   M.G.T. acknowledges fellowship support from the Engineering and Physical Sciences Research Council (EPSRC, UK).  The authors would also like to acknowledge helpful discussions will Will McCutcheon and David Barral.
\end{acknowledgments}

\appendix*

\section{The momentum operator}
\label{app:mo}
Here we give a brief outline of the use and derivation of the momentum operator.  When describing the evolution of a quantum state, we can choose to derive a time-dependent equation of motion using the Hamiltonian $\hat{H}(t)$, which is best suited to states confined within a cavity, or we can derive a spatially dependent equation of motion using the momentum operator, $\hat{M}(z)$, which is best suited to propagating fields.  Typically we might write the Hamiltonian as an integral over the energy density.  This density can be expressed in terms of wavevector or position, which are simply related to each other by a spatial Fourier transform:
\begin{equation}
\hat{H}(t) = \int \hat{\mathcal{H}}(\beta, t) d \beta = \int \hat{\mathcal{H}}(z,t) dz.
\end{equation}
Similarly, for the momentum we typically write this as an integral over the momentum density.  Again, this can be expressed as an integral over the angular frequency or time, both being related by Fourier transforms:
\begin{equation}
\hat{M}(z) = \int \hat{\mathcal{M}}(z,\omega) d \omega = \int \hat{\mathcal{M}} (z,t) d t.
\end{equation}
Which form is most convenient to start with depends on the particular problem to be solved, although all are of course physically equivalent.  In our work, we will consider electromagnetic fields propagating through a weakly-nonlinear material, so that the momentum can be described by,
\begin{equation}
\hat{M}(z) = \hbar \int \beta(\omega) \hat{a}^\dag (z,\omega) \hat{a}(z,\omega) d \omega,
\label{eqn:appM}
\end{equation}
where $\hbar \beta(\omega)$ is the momentum of a photon at frequency $\omega$ and $\hat{a}^\dag(z, \omega)\hat{a}(z,\omega)$ is the operator whose expectation value gives the total number-density of photons of frequency $\omega$ that cross a position $z$ in the time range $t \in ( -\infty, \infty )$ \cite{PhysRevA.42.4102}.  It is also convenient to introduce a slowly-varying spatio-temporal envelope operator $\hat{A}(z,t)$ such that $\hat{a}(z,t) = \hat{A}(z,t) \exp(i \beta_0 z - i \omega_0 t)$.  Then, we can see that,
\begin{equation}
\hat{a}(z,\omega) = \frac{1}{\sqrt{2 \pi}} e^{i \beta_0 z} \int \hat{A}(z,t) e^{i(\omega - \omega_0) t} d t.
\label{app:aA}
\end{equation}
Typically, the waveguides we use are short enough and bandwidth of the optical pulses sufficiently narrow that we can expand the dispersion relation up to first-order in frequency ($\beta(\omega) \approx \beta_0 + \frac{1}{v_g} (\omega-\omega_0)$).  We can then substitute this into (\ref{eqn:appM}) to find the momentum operator for the freely propagating signal and idler fields \cite{PhysRevLett.73.240}:
\begin{eqnarray}
\hat{M}_0(z) &=& \sum_{a=\{s,i\}} \hbar \int \beta_{0, a} \hat{A}_a^\dag\hat{A}_a + \nonumber \\
&& + \displaystyle \frac{i}{2 v_{g, a}}\left ( \hat{A}^\dag_a \frac{\partial \hat{A}_a}{\partial t} -\frac{\hat{A}^\dag_a}{\partial t} \hat{A}_a \right ) dt.
\label{app:mo}
\end{eqnarray}
The Heisenberg equation of motion for the spatial evolution of the free field envelope is given by:
\begin{equation}
\frac{d \hat{A}}{d z} = \frac{i}{\hbar} \left [ \hat{A}, \hat{M}_0 \right ] + \frac{\partial \hat{A}}{\partial z},
\end{equation}
where the partial derivative represents any explicity time-dependence of the operator given in its definition.  Substituting in the momentum operator above (\ref{app:mo}), we find the equation of motion for a narrow-bandwidth pulse propagating at the group velocity, as expected:
\begin{equation}
\frac{\partial \hat{A}}{\partial z} + \frac{1}{v_g} \frac{\partial \hat{A}}{\partial t} = 0.
\end{equation}
To include the nonlinearity we add a small nonlinear perturbation to the dispersion relation $\beta(\omega) \approx \beta_0 + \frac{1}{v_g} (\omega-\omega_0) + \Delta\beta$, \cite{agrawal2007nonlinear, mtp2}, where the perturbed wavevector is given by,
\begin{equation}
\Delta \beta = \frac{\omega_0}{c} \frac{\int F_0^*(x,y) \Delta n(x, y) F_0(x,y) dx dy}{\int \vert F_0(x,y) \vert^2 dx dy}.
\label{eqn:appDB}
\end{equation}
Here, we have assumed that the electric field is separable into transverse and longitudinal components, such that $E(x,y,z,t)=F(x,y)A(z,t) \exp \left (i \beta_0 z - i \omega_0 t \right )$ and $\Delta n(x,y)$ is the small nonlinear change in refractive index.  When calculating our perturbative results, we assume classical fields, but will ultimately make the identification $A(z,t) \rightarrow \hat{A}(z,t)$ for the signal and idler modes at the end of the derivation.  We assume that  three significant terms will give rise to small changes in refractive index.  For the pump field, only self-phase modulation will be of significance.  Then the nonlinear perturbation to the refractive index at the pump frequency is given by,
\begin{equation}
\Delta n_p(x,y) = n_2 \vert E_p(x,y) \vert^2.
\end{equation}
The signal and idler modes however, will experience cross-phase modulation and parametric amplification of the vacuum fluctations (spontaneous four-wave mixing) due to the strong pump.  The perturbation to the refractive index at the signal frequency will therefore be,
\begin{equation}
\Delta n_s(x,y) = 2 n_2 \vert E_p(x,y) \vert^2+ n_2 \frac{Ep^2 E_i^*}{E_s},
\end{equation}
with a similar expression for the idler.  These perturbations to the refractive index can be substituted into (\ref{eqn:appDB}) to find the corresponding perturbations to the wavevector at each frequency.  For the classical pump,we typically substitute the perturbed wavevector directly into the Nonlinear Schr\"{o}dinger Equation \cite{agrawal2007nonlinear} and find,
\begin{equation}
\frac{\partial A_p}{\partial z} + \frac{1}{v_g} \frac{\partial A_p}{\partial t} = i \gamma \vert A_p(z, t) \vert^2 A_p(z,t),
\end{equation}
where the nonlinear parameter is given by $\gamma = \frac{\omega_o n_2}{c A_{\text{eff}}}$  and it was assumed that the units of $\vert A(z,t) \vert^2$ are Watts.  Assuming that only the core of the waveguide is appreciably nonlinear, the effective area of the waveguide for this nonlinear interaction is found to be \cite{Lin:07},
\begin{equation}
A_{\text{eff}}=\frac{\vert \int \vert F_0(x,y) \vert^2 dx dy \vert^2}{\int_{\text{core}} \vert F_0(x,y) \vert^4 dx dy}.
\end{equation}
For the signal and idler fields, we wish to calculate the corresponding momentum operators.  To this end, we substitute the perturbed dispersion relations into (\ref{eqn:appM}) and express the result in the time-domain using (\ref{app:aA}).  Doing so, we find that the momentum operator that generates a cross-phase modulation on the signal and idler modes is expressed by,
\begin{equation}
\hat{M}_{XPM}(z) =2 \hbar \gamma \int \vert A_p(z,t) \vert^2 \hat{A}_a^\dag(z,t) \hat{A}_a(z,t) dt,
\end{equation}
where $a \in \{s, i \}$.  Finally, the momentum operator for spontaneous four-wave mixing on the signal and idler modes is found to be,
\begin{eqnarray}
\hat{M}_{SFWM}(z) &=& \hbar \gamma \exp \left [i \Delta \beta_0 z \right ] \nonumber \\
&&\times \int A_p^2(z,t) \hat{A}^\dag_i(z,t) \hat{A}^\dag_s(z,t).
\end{eqnarray}

% Create the reference section using BibTeX:
\bibliography{bibfile}

\end{document}